\documentclass[sigconf]{acmart}
\usepackage{graphicx}
\usepackage{float,subcaption}
\usepackage[capitalise]{cleveref}
\usepackage{amsmath}
\acmConference[DSHealth 2019]{KDD Workshop on Applied Data Science for
Healthcare}{August 2019} {Anchorage Alaska, United States}
\acmYear{2019}
\copyrightyear{2019}

\begin{document}
\title{Computational Phenotype Discovery\\via Probabilistic Independence}

\author{Thomas A. Lasko}
\affiliation{%
  \institution{Vanderbilt University Medical Center}
  \city{Nashville}
  \state{Tennessee}
  \country{USA}
}
\email{tom.lasko@vanderbilt.edu}

\author{Diego A. Mesa}
\affiliation{%
  \institution{Vanderbilt University Medical Center}
  \city{Nashville}
  \state{Tennessee}
  \country{USA}
}
\email{diego.mesa@vanderbilt.edu}

\begin{abstract}

Computational Phenotype Discovery research has taken various pragmatic
approaches to disentangling phenotypes from the episodic observations in
Electronic Health Records. In this work, we use transformation into continuous,
longitudinal curves to abstract away the sparse irregularity of the data, and we
introduce probabilistic independence as a guiding principle for disentangling
phenotypes into patterns that may more closely match true pathophysiologic
mechanisms. We use the identification of liver disease patterns that presage
development of Hepatocellular Carcinoma as a proof-of-concept demonstration.
  
\end{abstract}




\maketitle

\section{Introduction}
\label{sec:introduction}

There is growing evidence that insufficiently precise clinical diagnoses are
responsible for a large fraction of current treatment failures
\cite{Anderson2008, Ringman2014, Gutmann2014, Tuomi2014}. The goal of
Computational Phenotype Discovery is to create a more precise set of patterns
(or \emph{phenotypes}) of clinically observable variables that better represent
latent pathophysiologic mechanisms, and lead to more precise treatment decisions
\cite{Lasko2013}.

Electronic Health Records (EHRs) are a popular substrate for phenotype discovery
research, although the sparse, irregular, and asynchronous nature of the
observations they contain is a well-known barrier to secondary use
\cite{Lasko2013, Lasko2013b, Zhou2014, Ghassemi2015}.

A useful mental model is to consider the observations in a patient's record to
have been placed there by noisy episodic processes run within each of the
patient's medical conditions. But because those conditions share a common
vocabulary of observations, there is no obvious correspondence between the
observed data and the unobserved conditions. The problem of disentangling the
true latent processes from the sparse and irregularly observed data is the
phenotype discovery problem.

This problem maps well to unsupervised feature learning, and several approaches
have been taken to date. Some overcome sparsity and irregularity by aggregating
counts of events within defined time windows, producing one learning instance
per window. One line of research in this direction unraveled latent factors with
Nonnegative Matrix Factorization \cite{Ho2014a}. Another produced coarse
phenotypes (on the granularity of 'Cardiovascular Disease', or 'Lung Disease')
by decomposing the count matrix into a phenotype matrix and a densified temporal
expression matrix, using constraints of smoothness, sparsity, and nonnegativity
\cite{Zhou2014}.  A more ambitious effort used stacked denoising autoencoders to
decompose counts vectors for $700,000$ patients using a large set of counted
events into a set of $500$ phenotypes \cite{Miotto2016}.

Other researchers have used dense time-series data from an Intensive Care Unit
to avoid the episodic data problem. One group used denoising autoencoders as the
decomposition method \cite{Kale2015}, and another cast the problem as a
supervised multi-label problem, where clinical patterns were learned to predict
the assignment of a particular billing code \cite{Che2015}.

In this work we make two contributions. First, we overcome the episodic data
problem by transforming each sequence of repeated observations into a continuous
longitudinal curve. We demonstrate this transformation on the two most common
data types in an EHR. Second, we observe that while various methods have been
used to disentangle the latent phenotypes from the original basis of
observations, those methods appear to have been chosen from a pragmatic rather
than a principled perspective. We propose using \emph{probabilistic
independence} as a principled choice for that disentanglement because it is more
likely to illuminate actual disease mechanisms.

\section{Methods}
\label{sec:methods}

\subsection{Data}
\label{subsec:data}

All data for this project was extracted from the de-identified mirror of
Vanderbilt's Electronic Health Record, which contains administrative data,
billing codes, medication exposures, laboratory test results, and narrative text
for over 2 million patients, reaching back nearly 30 years \cite{Roden2008}. We
obtained IRB approval to use the data in this research.

From this source, we extracted all ICD-9 billing codes and all results from the
$500$ most common laboratory tests for all included records. We focused on
patients with liver disease, although we expected phenotypes from the full range
of conditions to appear among those patients.  Inclusion criteria for a patient
record were the presence of any lab result indicating at least mild liver
disease (AST > 40, ALT > 55, or Alk Phos > 150) or the presence of at least one
ICD-9 billing code indicating Nonalcoholic Fatty Liver Disease (571.8, 571.9, or
571.5). This retrieved $259,000$ records, from which $30,000$ were sampled at
random for tractability.  The ICD-9 codes were grouped into $1,814$
Phecodes\footnote{\url{https://phewascatalog.org/phecodes}}, for a total
instance dimension of $2,314$ variables.

\subsection{Longitudinal Curves}
\label{subsec:longitudinal-curves}

Previous work dealt with the episodic nature of medical data by aggregating that
data into observation windows, with the instance vector containing the count of
each variable's event within that window. That approach has an inherent
trade-off between statistical accuracy and window size. In contrast, we
converted each observation sequence into a continuous longitudinal curve, for
which any point on that curve estimates the instantaneous value for a given
measure of a given variable at that point.  The set of inferred curves can then
be sampled at any desired point, giving dense cross sections to use for
downstream learning. The semantics of each curve differ with the type of data it
represents.

Longitudinal curves for ICD-9 codes represent the intensity function for a
nonhomogeneous Gamma process over code-arrival events, and can be interpreted as
the instantaneous event density in time \cite{Lasko2014}. This is a powerful
representation that can account for burstiness, regularity, or randomness of
event timing, and can provide uncertainty distributions for the inferred curves.
Inferring a full model is computationally intensive, however, so for this work
we used an approximate method that produces a similar intensity curve, but
several orders of magnitude faster\footnote{
\url{https://github.com/ComputationalMedicineLab/fast_intensity}}.

Longitudinal curves for laboratory test results represent a distribution of
latent functions that could have produced the observed results. These were
originally computed using nonstationary Gaussian process regression
\cite{Lasko2015}. However, this inference is also computationally intensive, so
we approximated it using a smooth interpolation algorithm that provides a
similar mean curve, but without an uncertainty distribution, which is not
necessary for this work.

Other data types could be similarly represented by continuous curves, but these
two are the canonical types covering most of the structured data in a patient
record: event data with times and variable labels but no values, and
measurement data with times, variable labels, and measured values.

\begin{figure}[t!]
    \centering
    \includegraphics[width=\columnwidth]{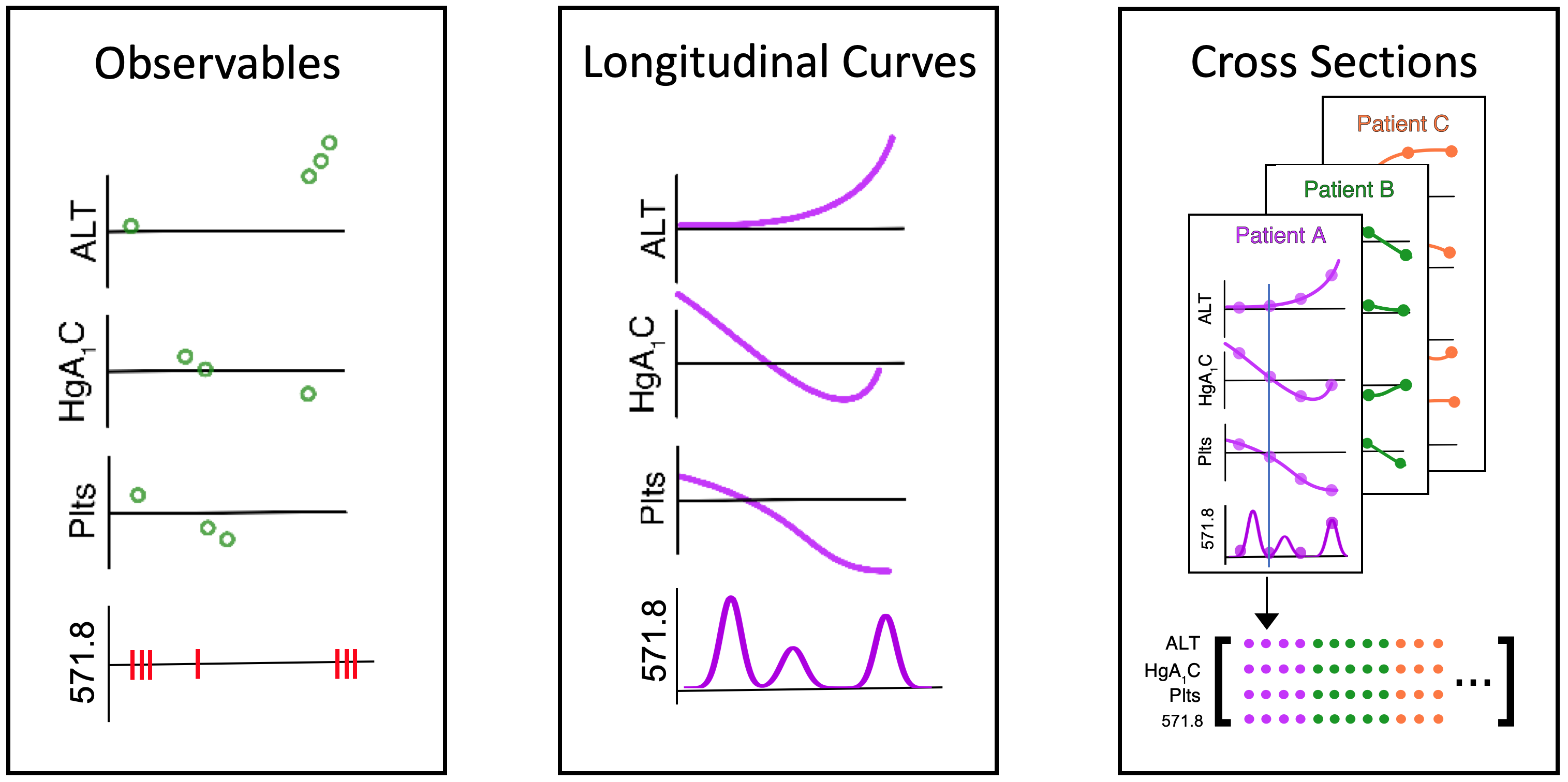}
    \caption{Preprocessing steps to produce the input dataset. Episodic data
      observations (left) are converted into continuous longitudinal curves
      (middle), which are then sampled at random points to produce cross
      sections (right), which form the dense data matrix (bottom right).}
    \Description[<short description>]{<long description>}
    \label{fig:method}
\end{figure}

\subsection{Independence}
\label{subsec:independence}

Our second contribution is the use of probabilistic independence as a principled
means for disentangling clinical phenotypes. Our underlying hypothesis is that
distinct pathophysiologic processes operate independently from each other, and
as such would leave probabilistically independent phenotypic fingerprints in the
record.

Independence has previously been proposed as a principle for disentangling
factors of variation in deep architectures
\cite{chen2018isolating,li2019disentangled}, but to our knowledge it has not
been described as a guiding principle for phenotype discovery. We anticipate
that deep architectures using this principle will eventually become the dominant
approach to phenotype discovery, but we do not use them in this work, because if
nothing else, the discovered phenotypes would be difficult to unambiguously
visualize and evaluate.

We chose Independent Components Analysis (ICA) \cite{Hyvarinen2001} as an
approach that isolates the principle of independence for disentangling
phenotypes and provides a linear decomposition that is easy to directly
visualize.  ICA also makes the assumption that phenotype compositions are
constant in time (although the level expressed by a given patient may change
with time). This is a useful simplification but not required by the domain
problem.

One instance of prior work does use ICA in passing as an un-optimized baseline
comparison, but does not directly evaluate the phenotypes produced
\cite{Miotto2016}.

\subsection{Experiment Details}
\label{subsec:experiment_details}

Records were extracted from the data source as described, and curves were
computed for each of the $n=2,314$ variables for each patient, for the duration
of each record. ICD-9 codes that did not occur in the record produced a constant
curve of zero intensity. Lab results that did not occur in the record were
filled a constant curve with the population median value. Cross sections of each
record's curve sets were then sampled at specific points in time chosen
uniformly at random at an average density of $1$ sample per year, giving
$m=180,369$ total instances (\cref{fig:method}).

This dataset $X \in \mathbb{R}^{n \times m}$ was then decomposed by ICA into $X
= AS$, where the phenotypic patterns of interest appeared in the columns of $A$,
and the patient-specific expression levels appeared in the rows of $S$. We
arbitrarily set the dimension of the phenotype space (the rank of $A$ and $S$)
to $500$.

\begin{figure*}[t!]
  \centering
  \begin{subfigure}[t]{0.33\textwidth}
    \centering
    \includegraphics[width=\textwidth]{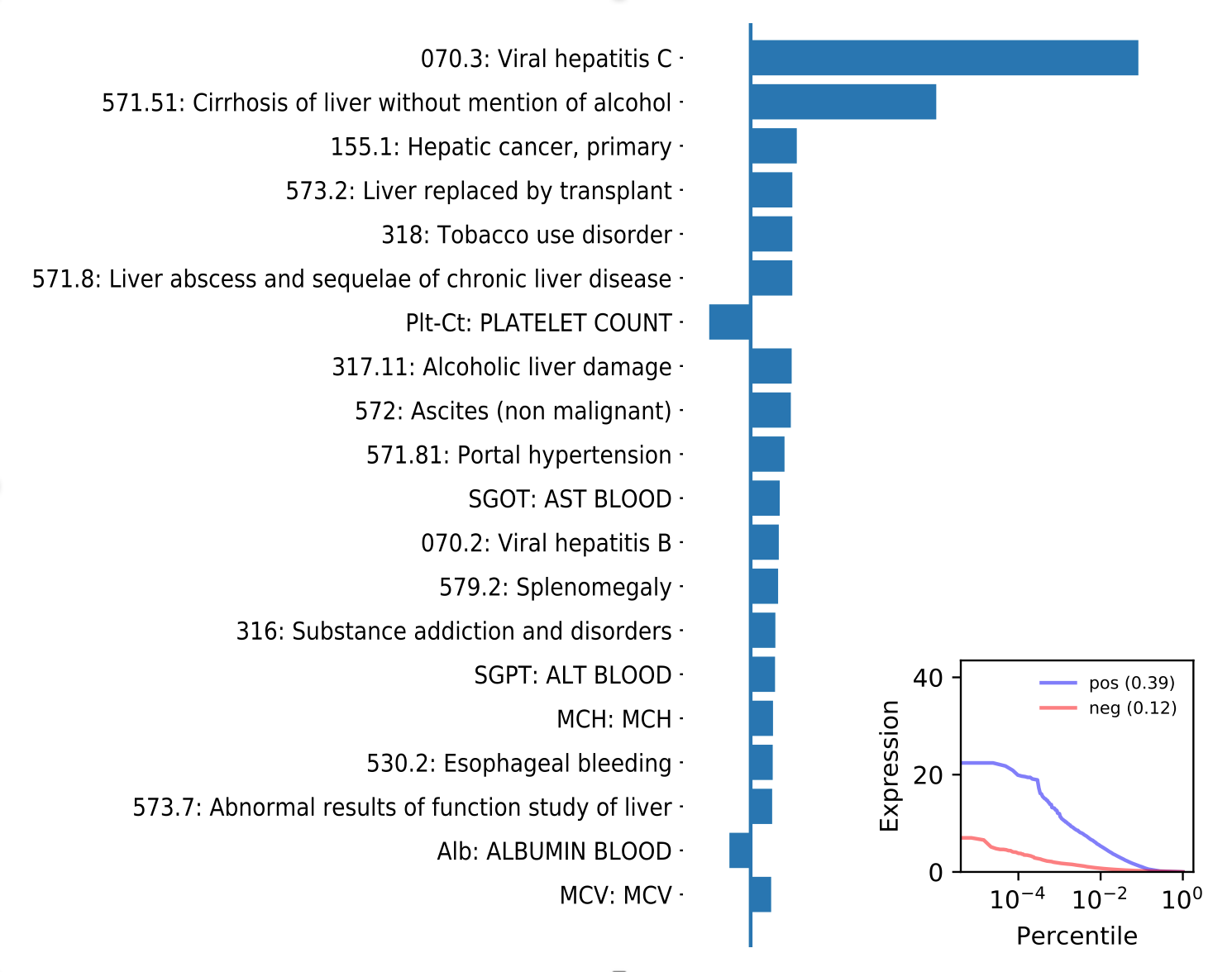}
    \caption{Late Hepatitis C}
    \label{fig:phenotypes-a}
  \end{subfigure}
  \begin{subfigure}[t]{0.33\textwidth}
    \centering
    \includegraphics[width=\textwidth]{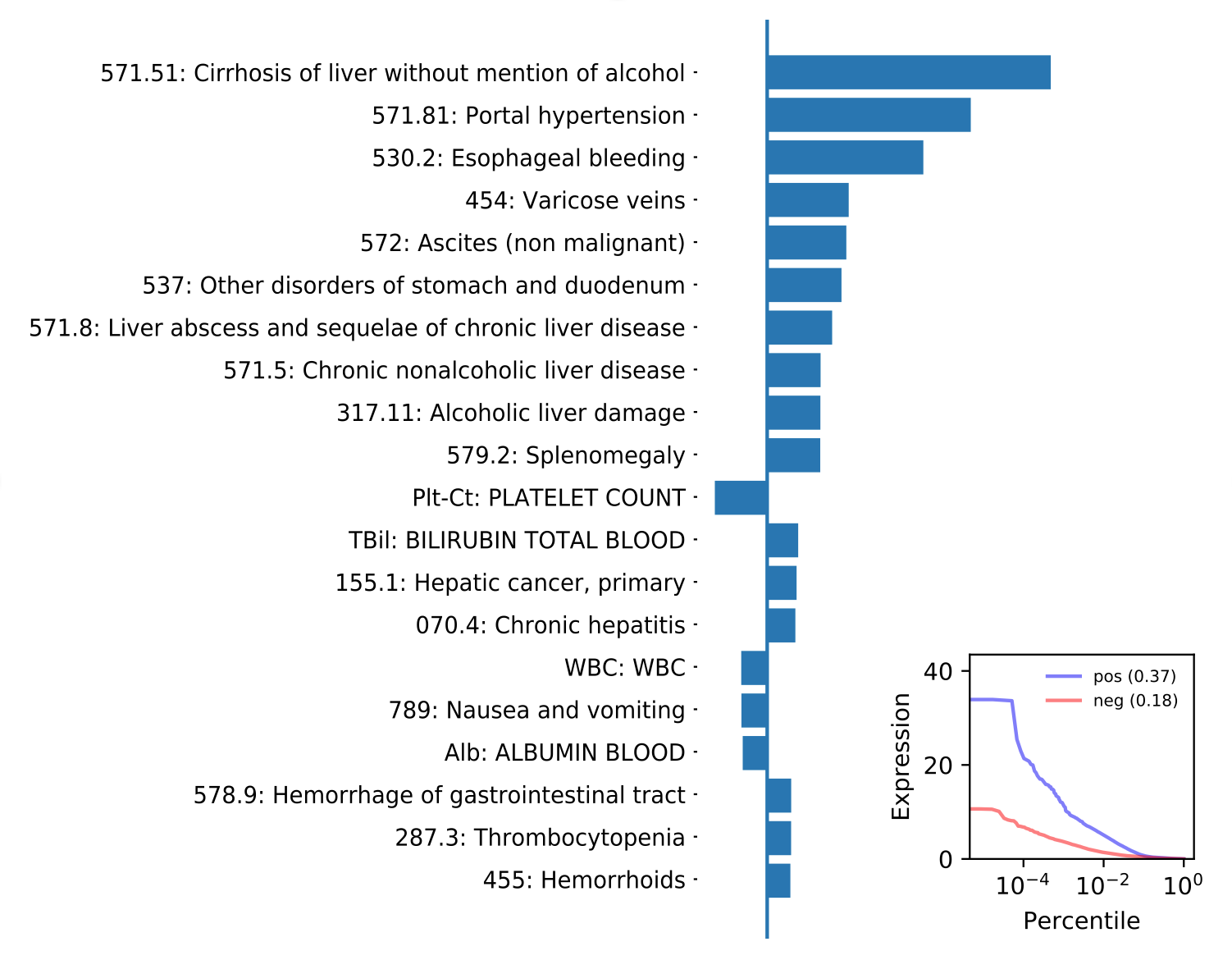}
    \caption{Early Nonalcoholic Cirrhosis}
    \label{fig:phenotypes-b}
  \end{subfigure}
  \begin{subfigure}[t]{0.33\textwidth}
    \centering
    \includegraphics[width=\textwidth]{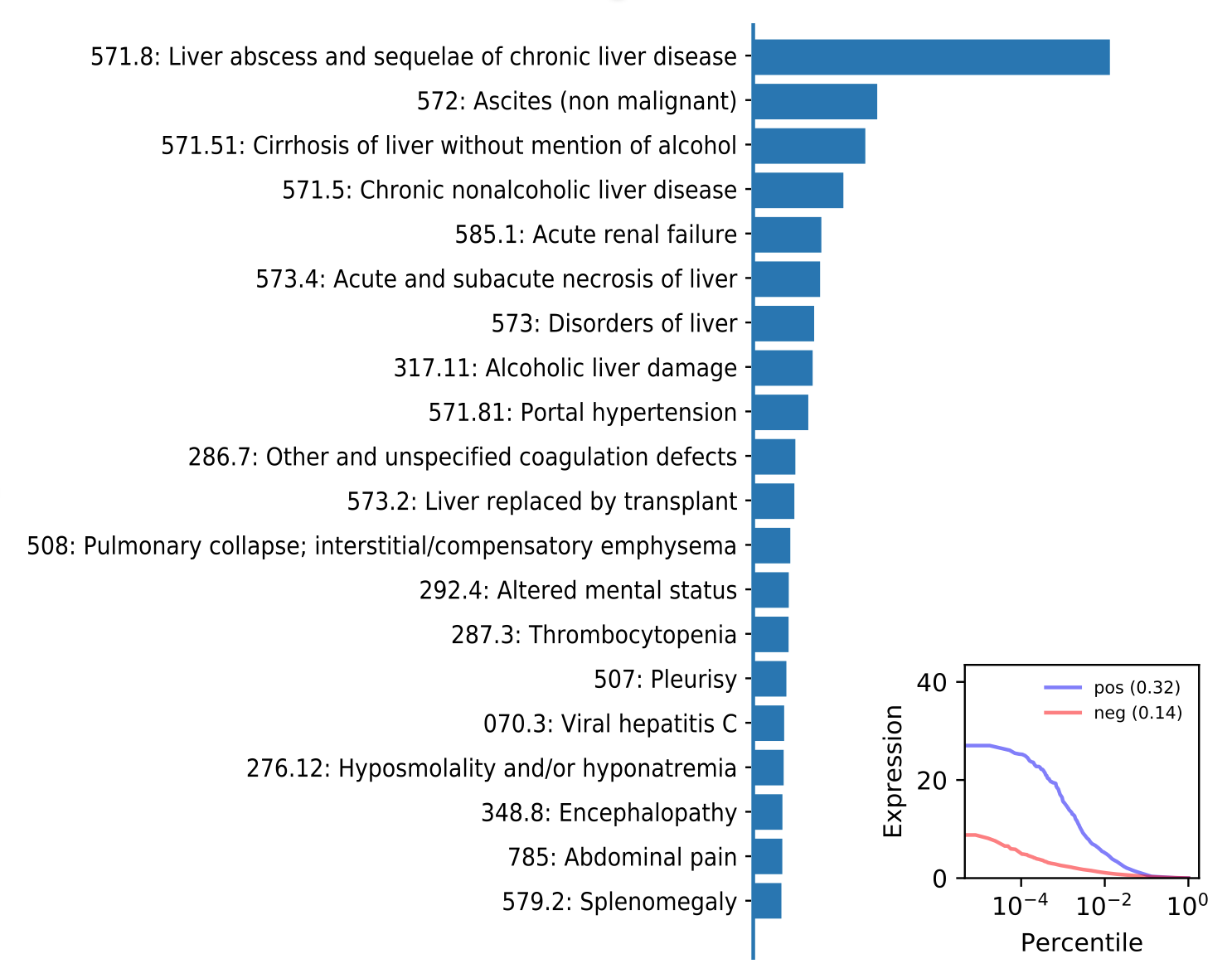}
    \caption{Late Nonalcoholic Cirrhosis}
    \label{fig:phenotypes-c}
  \end{subfigure}
  \vskip\baselineskip
  \begin{subfigure}[b]{0.33\textwidth}
    \centering
    \includegraphics[width=\textwidth]{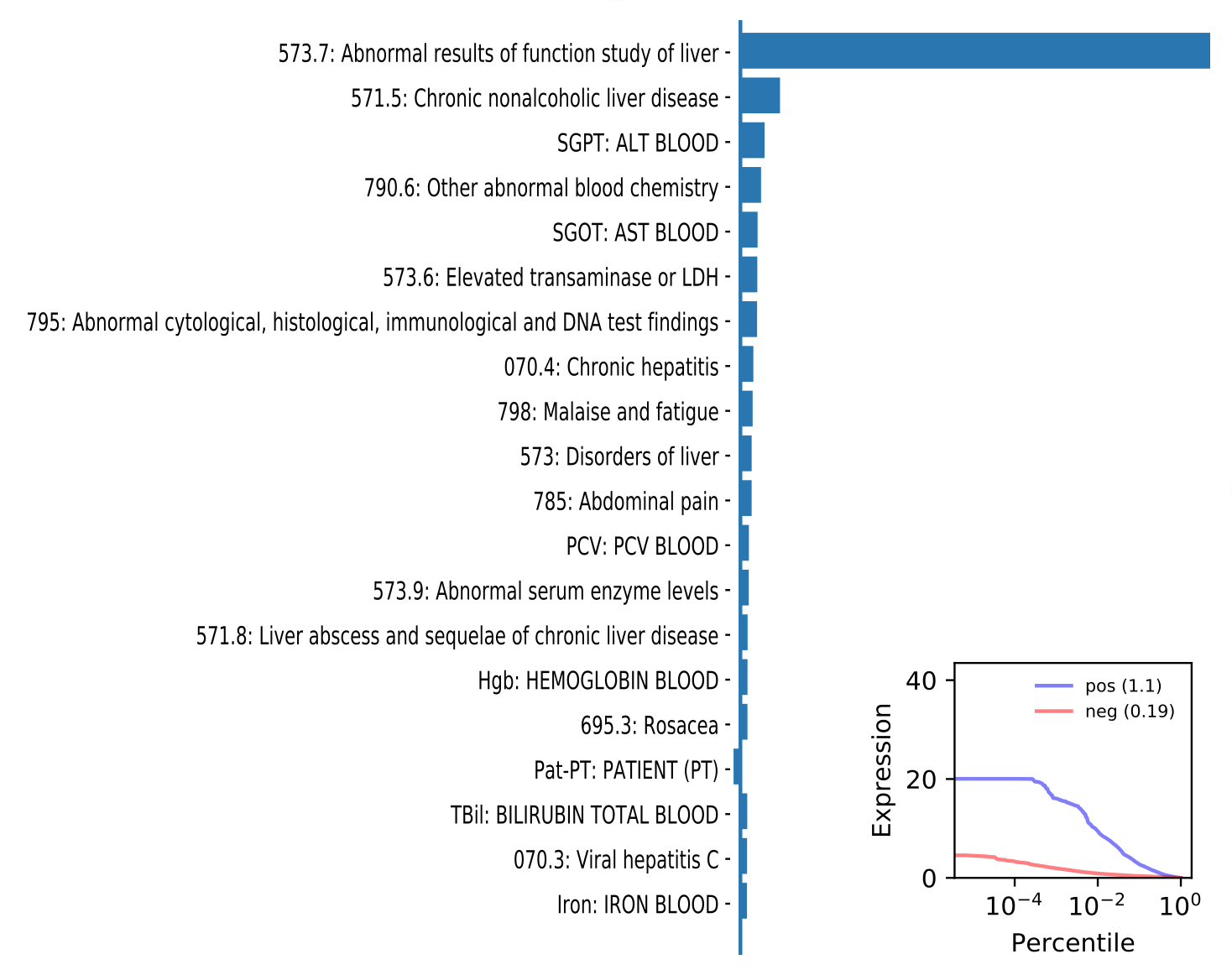}
    \caption{Mild Non-alcoholic Steatohepatitis}
    \label{fig:phenotypes-d}
  \end{subfigure}
  \begin{subfigure}[b]{0.33\textwidth}
    \centering
    \includegraphics[width=\textwidth]{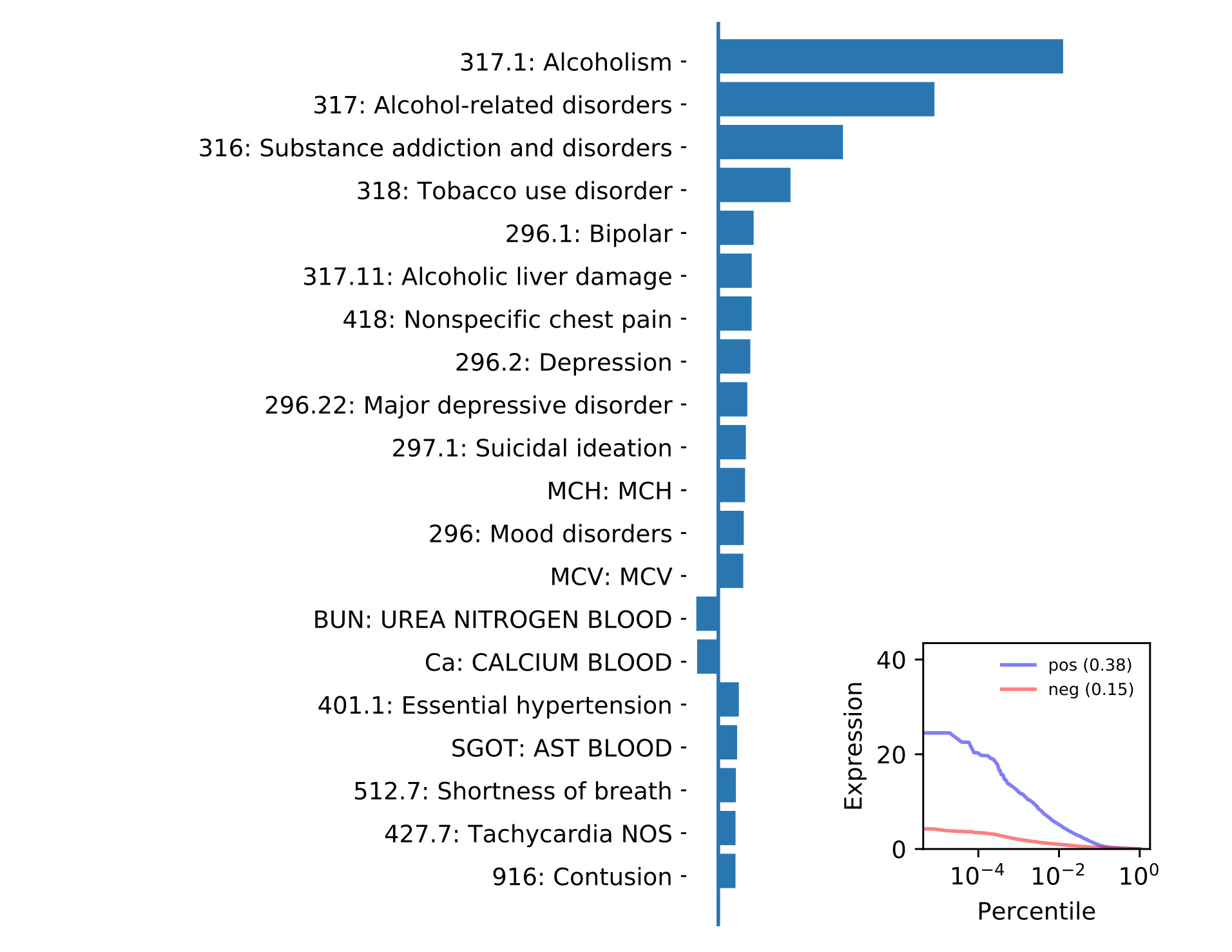}
    \caption{Alcohol Dependence}
    \label{fig:phenotypes-e}
  \end{subfigure}
  \begin{subfigure}[b]{0.33\textwidth}
    \centering
    \includegraphics[width=\textwidth]{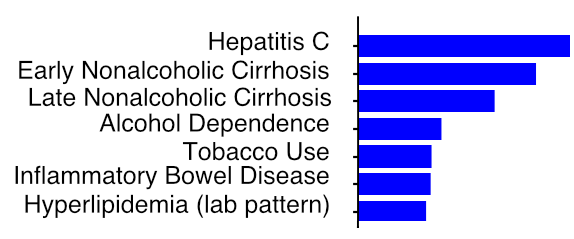}
    \vspace{5mm}
    \caption{HCC Precursors}
    \label{fig:phenotypes-f}
  \end{subfigure}
  \caption{Data-driven phenotypes include surprisingly detailed distinctions
    between early and late disease (a through e), and identify patterns that
    presage Hepatocellular Carcinoma (HCC) by 10 years (f). The disease name in
    each caption was assigned by an expert, not by the algorithm.}
  \Description[<short description>]{<long description>}
  \label{fig:phenotypes}
\end{figure*}

\subsection{Evaluation}
\label{subsec:evaluation}

The resulting phenotypes were evaluated in two ways: first, using clinical face
validity as judged by a domain expert, and second, by using them to investigate
conditions that presage the diagnosis of Hepatocellular Carcinoma (HCC, a type
of liver cancer).

The HCC investigation was formulated as a learning problem that predicts the
presence of an HCC billing code in the record exactly $10$ years from the time
of the cross section, given that there is no such code before then.  This is
different from (and harder than) the usual risk prediction problem, where the
goal would be to predict whether the event will occur any time over the next
$10$ years.

Our formulation allows us to examine the cancer disease process in terms of our
phenotypes a decade before its diagnosis. The salient piece of the results
(given adequate prediction accuracy) is an examination of the phenotypes that
are important to the prediction, and a judgment of whether they align with
what is known about the pathophysiology of HCC, and whether the results suggest
anything new about the disease.

We constructed negative instances from all records at least $10$ years long
that did not have an HCC diagnosis code ($155.0$, $155.1$, or $155.2$), and
positive instances from all those with at least $10$ years before the earliest
HCC code.  Records with fewer than $10$ pieces of data distributed among all
ICD codes and lab values were excluded. This produced $l=40,398$ instances, of
which $197$ were positive. We formed the raw data matrix
$X \in \mathbb{R}^{n \times l }$ as above, and then projected it into phenotype
space using $S = A^{-1}X$. A random forest was trained on the phenotype
expression matrix $S$ to predict the disease label.

\section{Results}
\label{sec:results}

The subjective evaluation of phenotype face validity was promising. About
$80-90\%$ of the discovered phenotypes were clearly recognizable as a specific
disease pattern. Among these were several phenotypes of mild to severe liver
disease at a surprisingly high level of differentiation
(\cref{fig:phenotypes}), such as very mild disease with only slight changes in
liver enzymes (panel d), early vs. late nonalcoholic cirrhosis patterns (panels
b and c), late-stage Hepatitis C (panel a), and alcohol dependence (panel e).
Of the non-recognizable phenotypes (not shown, for space) many described
complex laboratory result patterns with some underlying theme, but which did
not immediately suggest an obvious disease. We interpret these phenotypes as a
positive result, because they suggest laboratory patterns that may correspond
to specific disease mechanisms, but which have not yet been recognized, so
there are no billing codes for them.

The prediction model produced an adequate $0.76$ AUC, within the range of
historical results on the harder risk prediction problem.  The phenotype
forerunners of HCC, identified using the feature importance measure, are also
very satisfying. The conditions in our list are known to eventually produce the
tumor-generating environment needed for HCC, and their order here matches known
risk factors in the US \cite{Ma2016}. They also match exactly, and in order, the
top causes of liver transplant, which is the definitive treatment of HCC
\cite{Wong2015}. The next three phenotypes suggest interesting mechanisms that
reflect known or suspected risk factors \cite{Ma2016,
Koh2011,Rojas-Feria2013,Ioannou2016}.

\section{Discussion}
\label{sec:discussion}

This work introduces the use of independence for disentangling latent phenotypes
from EHR data, and it demonstrates the transformation of episodic clinical data
into continuous longitudinal curves to overcome well-known hurdles to learning
from that data.

Our approach learned $500$ different clinical patterns of billing codes and
laboratory values. These patterns provide a surprisingly granular set of
phenotypes that match clinical intuition, describing some conditions at multiple
stages, including subtle patterns of mild disease as well as obvious patterns of
severe disease. It is straightforward to extend the input data to include other
data types such as medications, demographics, and concepts in the text.

A predictive model used the learned phenotypes to identify a small set of
conditions that presage liver cancer by $10$ years, recovering known precursors
of that disease and strengthening previous suggestions about additional risk
factors. These findings suggest (but do not prove) a reasonable correspondence
to pathophysiologic mechanisms.

The number of phenotypes and the size of the input dataset are limited by
available memory under ICA decomposition, but are expandable by orders of
magnitude using deep architectures with suitable constraints to provide
independence. Deep architectures would also provide nonlinear decompositions,
but they would be more challenging to visualize and evaluate.

The best way of meaningfully evaluating the match between a set of data-driven
phenotypes and true pathophysiologic processes is currently an open question.
Previous work has commonly evaluated a set of phenotypes by their performance in
a clinical prediction model. We reject that approach because it doesn't measure
the most important property, which is how faithfully the phenotypes represent
actual disease mechanisms. Deep prediction model performance is a particularly
deceptive measure, because deep models are capable of making excellent
predictions from many different input representations, so long as those
representations somehow include sufficient information for the prediction.

In light of that open question, we evaluated our phenotypes using subjective
methods that try to assess how well the learned phenotypes correspond to what is
known about an \emph{a-priori} selected disease process, and whether they point
us in interesting clinical research directions. On both of these questions, our
learned phenotypes were quite promising.

\section*{Acknowledgements}
\label{sec:acknowledgements}

This work was funded in part by grant R01EB020666 from the National Institute of
Biomedical Imaging and Bioengineering, Vanderbilt's Academic Pathways
Fellowship, and a collaboration agreement with Pfizer, Inc. Clinical data was
provided by the Vanderbilt Synthetic Derivative, which is supported by
institutional funding and by the Vanderbilt CTSA grant ULTR000445.

\newcommand{\showDOI}[1]{\unskip}
\newcommand{\showURL}{\unskip}
\bibliographystyle{ACM-Reference-Format}
\bibliography{bibliography}


\begin{thebibliography}{23}

%%% the Web address.

\ifx \showCODEN    \undefined \def \showCODEN     #1{\unskip}     \fi
\ifx \showDOI      \undefined \def \showDOI       #1{#1}\fi
\ifx \showISBNx    \undefined \def \showISBNx     #1{\unskip}     \fi
\ifx \showISBNxiii \undefined \def \showISBNxiii  #1{\unskip}     \fi
\ifx \showISSN     \undefined \def \showISSN      #1{\unskip}     \fi
\ifx \showLCCN     \undefined \def \showLCCN      #1{\unskip}     \fi
\ifx \shownote     \undefined \def \shownote      #1{#1}          \fi
\ifx \showarticletitle \undefined \def \showarticletitle #1{#1}   \fi
\ifx \showURL      \undefined \def \showURL       {\relax}        \fi
% invisible to TeX
\providecommand\bibfield[2]{#2}
\providecommand\bibinfo[2]{#2}
\providecommand\natexlab[1]{#1}
\providecommand\showeprint[2][]{arXiv:#2}

\bibitem[\protect\citeauthoryear{Anderson}{Anderson}{2008}]%
        {Anderson2008}
\bibfield{author}{\bibinfo{person}{Gary~P. Anderson}.}
  \bibinfo{year}{2008}\natexlab{}.
\newblock \showarticletitle{Endotyping asthma: new insights into key pathogenic
  mechanisms in a complex, heterogeneous disease.}
\newblock \bibinfo{journal}{\emph{Lancet}} \bibinfo{volume}{372},
  \bibinfo{number}{9643} (\bibinfo{date}{Sep} \bibinfo{year}{2008}),
  \bibinfo{pages}{1107--1119}.
\newblock
\urldef\tempurl%
\url{https://doi.org/10.1016/S0140-6736(08)61452-X}
\showDOI{\tempurl}


\bibitem[\protect\citeauthoryear{Che, Kale, Li, Bahadori, and Liu}{Che
  et~al\mbox{.}}{2015}]%
        {Che2015}
\bibfield{author}{\bibinfo{person}{Zhengpoing Che}, \bibinfo{person}{David
  Kale}, \bibinfo{person}{Wenzhe Li}, \bibinfo{person}{Mohammad~Taha Bahadori},
  {and} \bibinfo{person}{Yan Liu}.} \bibinfo{year}{2015}\natexlab{}.
\newblock \showarticletitle{Deep Computational Phenotyping}. In
  \bibinfo{booktitle}{\emph{Proceedings of the 21th ACM SIGKDD International
  Conference on Knowledge Discovery and Data Mining (KDD'15)}}.
\newblock


\bibitem[\protect\citeauthoryear{Chen, Li, Grosse, and Duvenaud}{Chen
  et~al\mbox{.}}{2018}]%
        {chen2018isolating}
\bibfield{author}{\bibinfo{person}{Tian~Qi Chen}, \bibinfo{person}{Xuechen Li},
  \bibinfo{person}{Roger~B Grosse}, {and} \bibinfo{person}{David~K Duvenaud}.}
  \bibinfo{year}{2018}\natexlab{}.
\newblock \showarticletitle{Isolating sources of disentanglement in variational
  autoencoders}. In \bibinfo{booktitle}{\emph{NIPS 2018}}.
  \bibinfo{pages}{2610--2620}.
\newblock


\bibitem[\protect\citeauthoryear{Ghassemi, Pimentel, Naumann, Brennan, Clifton,
  Szolovits, and Feng}{Ghassemi et~al\mbox{.}}{2015}]%
        {Ghassemi2015}
\bibfield{author}{\bibinfo{person}{Marzyeh Ghassemi}, \bibinfo{person}{Marco
  A~F Pimentel}, \bibinfo{person}{Tristan Naumann}, \bibinfo{person}{Thomas
  Brennan}, \bibinfo{person}{David~A Clifton}, \bibinfo{person}{Peter
  Szolovits}, {and} \bibinfo{person}{Mengling Feng}.}
  \bibinfo{year}{2015}\natexlab{}.
\newblock \showarticletitle{A Multivariate Timeseries Modeling Approach to
  Severity of Illness Assessment and Forecasting in ICU with Sparse,
  Heterogeneous Clinical Data.}
\newblock \bibinfo{journal}{\emph{AAAI 2015}}  \bibinfo{volume}{2015}
  (\bibinfo{date}{Jan.} \bibinfo{year}{2015}), \bibinfo{pages}{446--453}.
\newblock
\showISSN{2159-5399}


\bibitem[\protect\citeauthoryear{Gutmann}{Gutmann}{2014}]%
        {Gutmann2014}
\bibfield{author}{\bibinfo{person}{David~H. Gutmann}.}
  \bibinfo{year}{2014}\natexlab{}.
\newblock \showarticletitle{Eliminating barriers to personalized medicine:
  Learning from neurofibromatosis type 1.}
\newblock \bibinfo{journal}{\emph{Neurology}} (\bibinfo{date}{Jun}
  \bibinfo{year}{2014}).
\newblock
\urldef\tempurl%
\url{https://doi.org/10.1212/WNL.0000000000000652}
\showDOI{\tempurl}


\bibitem[\protect\citeauthoryear{Ho, Ghosh, and Sun}{Ho et~al\mbox{.}}{2014}]%
        {Ho2014a}
\bibfield{author}{\bibinfo{person}{Joyce~C. Ho}, \bibinfo{person}{Joydeep
  Ghosh}, {and} \bibinfo{person}{Jimeng Sun}.} \bibinfo{year}{2014}\natexlab{}.
\newblock \showarticletitle{Marble: High-throughput Phenotyping from Electronic
  Health Records via Sparse Nonnegative Tensor Factorization}. In
  \bibinfo{booktitle}{\emph{KDD 2014}} \emph{(\bibinfo{series}{KDD '14})}.
  \bibinfo{publisher}{ACM}, \bibinfo{address}{New York, NY, USA},
  \bibinfo{pages}{115--124}.
\newblock
\showISBNx{978-1-4503-2956-9}
\urldef\tempurl%
\url{https://doi.org/10.1145/2623330.2623658}
\showDOI{\tempurl}


\bibitem[\protect\citeauthoryear{Hyvarinen, Karhunen, and Oja}{Hyvarinen
  et~al\mbox{.}}{2001}]%
        {Hyvarinen2001}
\bibfield{author}{\bibinfo{person}{Aapo Hyvarinen}, \bibinfo{person}{Juha
  Karhunen}, {and} \bibinfo{person}{Erkki Oja}.}
  \bibinfo{year}{2001}\natexlab{}.
\newblock \bibinfo{booktitle}{\emph{Independent Component Analysis}}.
\newblock \bibinfo{publisher}{Wiley}, \bibinfo{address}{New York}.
\newblock


\bibitem[\protect\citeauthoryear{Ioannou}{Ioannou}{2016}]%
        {Ioannou2016}
\bibfield{author}{\bibinfo{person}{George~N Ioannou}.}
  \bibinfo{year}{2016}\natexlab{}.
\newblock \showarticletitle{The Role of Cholesterol in the Pathogenesis of
  NASH.}
\newblock \bibinfo{journal}{\emph{Trends Endocrinol Metab}}
  \bibinfo{volume}{27} (\bibinfo{date}{Feb.} \bibinfo{year}{2016}),
  \bibinfo{pages}{84--95}.
\newblock
Issue 2.
\showISSN{1879-3061}
\urldef\tempurl%
\url{https://doi.org/10.1016/j.tem.2015.11.008}
\showDOI{\tempurl}


\bibitem[\protect\citeauthoryear{Kale, Che, Bahadori, Li, Liu, and Wetzel}{Kale
  et~al\mbox{.}}{2015}]%
        {Kale2015}
\bibfield{author}{\bibinfo{person}{David~C. Kale}, \bibinfo{person}{Zhengping
  Che}, \bibinfo{person}{Mohammad~Taha Bahadori}, \bibinfo{person}{Wenzhe Li},
  \bibinfo{person}{Yan Liu}, {and} \bibinfo{person}{Randall Wetzel}.}
  \bibinfo{year}{2015}\natexlab{}.
\newblock \showarticletitle{Causal Phenotype Discovery via Deep Networks}. In
  \bibinfo{booktitle}{\emph{Proceedings AMIA Symposium 2015}}.
\newblock


\bibitem[\protect\citeauthoryear{Koh, Robien, Wang, Govindarajan, Yuan, and
  Yu}{Koh et~al\mbox{.}}{2011}]%
        {Koh2011}
\bibfield{author}{\bibinfo{person}{W-P Koh}, \bibinfo{person}{K Robien},
  \bibinfo{person}{R Wang}, \bibinfo{person}{S Govindarajan},
  \bibinfo{person}{J-M Yuan}, {and} \bibinfo{person}{M~C Yu}.}
  \bibinfo{year}{2011}\natexlab{}.
\newblock \showarticletitle{Smoking as an independent risk factor for
  hepatocellular carcinoma: the Singapore Chinese Health Study.}
\newblock \bibinfo{journal}{\emph{Br J Cancer}}  \bibinfo{volume}{105}
  (\bibinfo{date}{Oct.} \bibinfo{year}{2011}), \bibinfo{pages}{1430--1435}.
\newblock
Issue 9.
\showISSN{1532-1827}
\urldef\tempurl%
\url{https://doi.org/10.1038/bjc.2011.360}
\showDOI{\tempurl}


\bibitem[\protect\citeauthoryear{Lasko}{Lasko}{2013}]%
        {Lasko2013b}
\bibfield{author}{\bibinfo{person}{Thomas~A Lasko}.}
  \bibinfo{year}{2013}\natexlab{}.
\newblock \showarticletitle{Inferring the Latent Intensity of Clinical Events
  Using Modulated Renewal Processes}. In \bibinfo{booktitle}{\emph{NIPS 2013
  Workshop on Machine Learning for Clinical Data Analysis and Healthcare}}.
\newblock


\bibitem[\protect\citeauthoryear{Lasko}{Lasko}{2014}]%
        {Lasko2014}
\bibfield{author}{\bibinfo{person}{Thomas~A. Lasko}.}
  \bibinfo{year}{2014}\natexlab{}.
\newblock \showarticletitle{Efficient Inference of {G}aussian Process Modulated
  Renewal Processes with Application to Medical Event Data}. In
  \bibinfo{booktitle}{\emph{Proceedings of the Thirtieth Conference on
  Uncertainty in Artificial Intelligence {(UAI)}}}.
\newblock
\showeprint{1402.4732}


\bibitem[\protect\citeauthoryear{Lasko}{Lasko}{2015}]%
        {Lasko2015}
\bibfield{author}{\bibinfo{person}{Thomas~A Lasko}.}
  \bibinfo{year}{2015}\natexlab{}.
\newblock \showarticletitle{Nonstationary Gaussian Process Regression for
  Evaluating Clinical Laboratory Test Sampling Strategies}.
\newblock \bibinfo{journal}{\emph{AAAI 2015}} (\bibinfo{date}{Jan}
  \bibinfo{year}{2015}), \bibinfo{pages}{1777--1783}.
\newblock
\showISSN{2159-5399}


\bibitem[\protect\citeauthoryear{Lasko, Denny, and Levy}{Lasko
  et~al\mbox{.}}{2013}]%
        {Lasko2013}
\bibfield{author}{\bibinfo{person}{Thomas~A. Lasko}, \bibinfo{person}{Joshua~C.
  Denny}, {and} \bibinfo{person}{Mia~A. Levy}.}
  \bibinfo{year}{2013}\natexlab{}.
\newblock \showarticletitle{Computational Phenotype Discovery Using
  Unsupervised Feature Learning over Noisy, Sparse, and Irregular Clinical
  Data.}
\newblock \bibinfo{journal}{\emph{PLoS One}} \bibinfo{volume}{8},
  \bibinfo{number}{6} (\bibinfo{year}{2013}), \bibinfo{pages}{e66341}.
\newblock
\urldef\tempurl%
\url{https://doi.org/10.1371/journal.pone.0066341}
\showDOI{\tempurl}


\bibitem[\protect\citeauthoryear{Li, Pan, Wang, Peng, Yang, and Cambria}{Li
  et~al\mbox{.}}{2019}]%
        {li2019disentangled}
\bibfield{author}{\bibinfo{person}{Yang Li}, \bibinfo{person}{Quan Pan},
  \bibinfo{person}{Suhang Wang}, \bibinfo{person}{Haiyun Peng},
  \bibinfo{person}{Tao Yang}, {and} \bibinfo{person}{Erik Cambria}.}
  \bibinfo{year}{2019}\natexlab{}.
\newblock \showarticletitle{Disentangled variational auto-encoder for
  semi-supervised learning}.
\newblock \bibinfo{journal}{\emph{Information Sciences}}  \bibinfo{volume}{482}
  (\bibinfo{year}{2019}), \bibinfo{pages}{73--85}.
\newblock


\bibitem[\protect\citeauthoryear{Ma, Yang, Tu, Gao, Tan, Zheng, Bray, and
  Xiang}{Ma et~al\mbox{.}}{2016}]%
        {Ma2016}
\bibfield{author}{\bibinfo{person}{Xiao Ma}, \bibinfo{person}{Yang Yang},
  \bibinfo{person}{Hong Tu}, \bibinfo{person}{Jing Gao},
  \bibinfo{person}{Yu-Ting Tan}, \bibinfo{person}{Jia-Li Zheng},
  \bibinfo{person}{Freddie Bray}, {and} \bibinfo{person}{Yong-Bing Xiang}.}
  \bibinfo{year}{2016}\natexlab{}.
\newblock \showarticletitle{Risk prediction models for hepatocellular carcinoma
  in different populations.}
\newblock \bibinfo{journal}{\emph{Chin J Cancer Res}}  \bibinfo{volume}{28}
  (\bibinfo{date}{April} \bibinfo{year}{2016}), \bibinfo{pages}{150--160}.
\newblock
Issue 2.
\showISSN{1000-9604}
\urldef\tempurl%
\url{https://doi.org/10.21147/j.issn.1000-9604.2016.02.02}
\showDOI{\tempurl}


\bibitem[\protect\citeauthoryear{Miotto, Li, Kidd, and Dudley}{Miotto
  et~al\mbox{.}}{2016}]%
        {Miotto2016}
\bibfield{author}{\bibinfo{person}{Riccardo Miotto}, \bibinfo{person}{Li Li},
  \bibinfo{person}{Brian~A. Kidd}, {and} \bibinfo{person}{Joel~T. Dudley}.}
  \bibinfo{year}{2016}\natexlab{}.
\newblock \showarticletitle{Deep Patient: An Unsupervised Representation to
  Predict the Future of Patients from the Electronic Health Records.}
\newblock \bibinfo{journal}{\emph{Sci Rep}}  \bibinfo{volume}{6}
  (\bibinfo{year}{2016}), \bibinfo{pages}{26094}.
\newblock
\urldef\tempurl%
\url{https://doi.org/10.1038/srep26094}
\showDOI{\tempurl}


\bibitem[\protect\citeauthoryear{Ringman, Goate, Masters, Cairns, Danek,
  Graff-Radford, Ghetti, and Morris}{Ringman et~al\mbox{.}}{2014}]%
        {Ringman2014}
\bibfield{author}{\bibinfo{person}{JohnM. Ringman}, \bibinfo{person}{Alison
  Goate}, \bibinfo{person}{ColinL. Masters}, \bibinfo{person}{NigelJ. Cairns},
  \bibinfo{person}{Adrian Danek}, \bibinfo{person}{Neill Graff-Radford},
  \bibinfo{person}{Bernardino Ghetti}, {and} \bibinfo{person}{JohnC. Morris}.}
  \bibinfo{year}{2014}\natexlab{}.
\newblock \showarticletitle{Genetic Heterogeneity in Alzheimer Disease and
  Implications for Treatment Strategies}.
\newblock \bibinfo{journal}{\emph{Curr Neurol Neurosci Rep}}
  \bibinfo{volume}{14}, \bibinfo{number}{11}, Article \bibinfo{articleno}{499}
  (\bibinfo{year}{2014}).
\newblock
\showISSN{1528-4042}
\urldef\tempurl%
\url{https://doi.org/10.1007/s11910-014-0499-8}
\showDOI{\tempurl}


\bibitem[\protect\citeauthoryear{Roden, Pulley, Basford, Bernard, Clayton,
  Balser, and Masys}{Roden et~al\mbox{.}}{2008}]%
        {Roden2008}
\bibfield{author}{\bibinfo{person}{D.~M. Roden}, \bibinfo{person}{J.~M.
  Pulley}, \bibinfo{person}{M.~A. Basford}, \bibinfo{person}{G.~R. Bernard},
  \bibinfo{person}{E.~W. Clayton}, \bibinfo{person}{J.~R. Balser}, {and}
  \bibinfo{person}{D.~R. Masys}.} \bibinfo{year}{2008}\natexlab{}.
\newblock \showarticletitle{{{D}evelopment of a large-scale de-identified
  {D}{N}{A} biobank to enable personalized medicine}}.
\newblock \bibinfo{journal}{\emph{Clin Pharmacol Ther}} \bibinfo{volume}{84},
  \bibinfo{number}{3} (\bibinfo{date}{Sep} \bibinfo{year}{2008}),
  \bibinfo{pages}{362--369}.
\newblock


\bibitem[\protect\citeauthoryear{Rojas-Feria, Castro, Su{\'a}rez, Ampuero, and
  Romero-G{\'o}mez}{Rojas-Feria et~al\mbox{.}}{2013}]%
        {Rojas-Feria2013}
\bibfield{author}{\bibinfo{person}{Mar{\'\i}a Rojas-Feria},
  \bibinfo{person}{Manuel Castro}, \bibinfo{person}{Emilio Su{\'a}rez},
  \bibinfo{person}{Javier Ampuero}, {and} \bibinfo{person}{Manuel
  Romero-G{\'o}mez}.} \bibinfo{year}{2013}\natexlab{}.
\newblock \showarticletitle{Hepatobiliary manifestations in inflammatory bowel
  disease: the gut, the drugs and the liver.}
\newblock \bibinfo{journal}{\emph{World J Gastroenterol}}  \bibinfo{volume}{19}
  (\bibinfo{date}{Nov.} \bibinfo{year}{2013}), \bibinfo{pages}{7327--7340}.
\newblock
Issue 42.
\showISSN{2219-2840}
\urldef\tempurl%
\url{https://doi.org/10.3748/wjg.v19.i42.7327}
\showDOI{\tempurl}


\bibitem[\protect\citeauthoryear{Tuomi, Santoro, Caprio, Cai, Weng, and
  Groop}{Tuomi et~al\mbox{.}}{2014}]%
        {Tuomi2014}
\bibfield{author}{\bibinfo{person}{Tiinamaija Tuomi}, \bibinfo{person}{Nicola
  Santoro}, \bibinfo{person}{Sonia Caprio}, \bibinfo{person}{Mengyin Cai},
  \bibinfo{person}{Jianping Weng}, {and} \bibinfo{person}{Leif Groop}.}
  \bibinfo{year}{2014}\natexlab{}.
\newblock \showarticletitle{The many faces of diabetes: a disease with
  increasing heterogeneity}.
\newblock \bibinfo{journal}{\emph{Lancet}} \bibinfo{volume}{383},
  \bibinfo{number}{9922} (\bibinfo{year}{2014}), \bibinfo{pages}{1084--1094}.
\newblock
\showISSN{0140-6736}
\urldef\tempurl%
\url{https://doi.org/10.1016/S0140-6736(13)62219-9}
\showDOI{\tempurl}


\bibitem[\protect\citeauthoryear{Wong, Aguilar, Cheung, Perumpail, Harrison,
  Younossi, and Ahmed}{Wong et~al\mbox{.}}{2015}]%
        {Wong2015}
\bibfield{author}{\bibinfo{person}{Robert~J Wong}, \bibinfo{person}{Maria
  Aguilar}, \bibinfo{person}{Ramsey Cheung}, \bibinfo{person}{Ryan~B
  Perumpail}, \bibinfo{person}{Stephen~A Harrison}, \bibinfo{person}{Zobair~M
  Younossi}, {and} \bibinfo{person}{Aijaz Ahmed}.}
  \bibinfo{year}{2015}\natexlab{}.
\newblock \showarticletitle{Nonalcoholic steatohepatitis is the second leading
  etiology of liver disease among adults awaiting liver transplantation in the
  United States.}
\newblock \bibinfo{journal}{\emph{Gastroenterology}}  \bibinfo{volume}{148}
  (\bibinfo{date}{March} \bibinfo{year}{2015}), \bibinfo{pages}{547--555}.
\newblock
Issue 3.
\showISSN{1528-0012}
\urldef\tempurl%
\url{https://doi.org/10.1053/j.gastro.2014.11.039}
\showDOI{\tempurl}


\bibitem[\protect\citeauthoryear{Zhou, Wang, Hu, and Ye}{Zhou
  et~al\mbox{.}}{2014}]%
        {Zhou2014}
\bibfield{author}{\bibinfo{person}{Jiayu Zhou}, \bibinfo{person}{Fei Wang},
  \bibinfo{person}{Jianying Hu}, {and} \bibinfo{person}{Jieping Ye}.}
  \bibinfo{year}{2014}\natexlab{}.
\newblock \showarticletitle{From Micro to Macro: Data Driven Phenotyping by
  Densification of Longitudinal Electronic Medical Records}. In
  \bibinfo{booktitle}{\emph{KDD 2014}} \emph{(\bibinfo{series}{KDD '14})}.
  \bibinfo{publisher}{ACM}, \bibinfo{address}{New York, NY, USA},
  \bibinfo{pages}{135--144}.
\newblock
\showISBNx{978-1-4503-2956-9}
\urldef\tempurl%
\url{https://doi.org/10.1145/2623330.2623711}
\showDOI{\tempurl}


\end{thebibliography}

\end{document}